\documentclass[preprint]{aastex}
\usepackage{psfig}
\newcommand{\lta}{$\; \buildrel < \over \sim \;$}
\newcommand{\simlt}{\lower.5ex\hbox{\lta}}
\newcommand{\gta}{$\; \buildrel > \over \sim \;$}
\newcommand{\simgt}{\lower.5ex\hbox{\gta}}

\slugcomment{Accepted to the P.A.S.P.}
\shortauthors{Belle, Howell, \& Mills}
\shorttitle{QQ Vul}

\begin{document}

\title{{\it EUVE} Observations of the Magnetic Cataclysmic Variable 
QQ~Vulpeculae}

\author{Kunegunda E. Belle, Steve B. Howell\altaffilmark{1} 
\& Amy Mills}
\affil{Department of Physics and Astronomy, University of Wyoming,
Laramie, WY 82071; \\ keb@tana.irastro.uwyo.edu; howell@psi.edu}

\altaffiltext{1}{
Present address: Head, Astrophysics Group, Planetary Science
Institute, 620 N. 6th Ave.  Tucson, AZ 85705}

\begin{abstract}
We present simultaneous X-ray ($\lambda_{\rm peak} \sim 44{\rm \AA}$)
and EUV ($\lambda_{\rm peak} = 89{\rm \AA}$) light curves for the
magnetic cataclysmic variable QQ~Vulpeculae, obtained with the {\it
EUVE} satellite.  We find that the unique shape of the X-ray light
curve is different from previously obtained X-ray light curves of
QQ~Vul and provides evidence for two-pole accretion.  Detailed
examination of the photometric data indicates that QQ~Vul undergoes a
stellar eclipse of the X-ray emitting region, indicative of a high
binary inclination.  We discuss possible
implications for the nature of this system given the observed shape of
its EUV and X-ray light curves.
\end{abstract}

\keywords{accretion --- cataclysmic variables --- 
stars: individual (QQ Vul) --- ultraviolet: stars --- X-rays: stars}

\section{Introduction}\label{intro}
Polars, or AM Her systems, a subset of magnetic cataclysmic variables
(CVs), are a unique class of CVs.  As with the typical CV, polars
contain a white dwarf which accretes matter from a Roche lobe-filling
low-mass secondary star.  However, unlike a typical CV, the white
dwarf in a polar has a magnetic field strength on the order of a few
tens of megaGauss (MG).  These strong magnetic fields cause the
disruption of the accreted material that would otherwise form an
accretion disk.  Instead, material transfer is routed on to the white
dwarf in the form of an accretion stream which follows the path of one
or more of the magnetic field lines to impact the surface at one or
both of the magnetic poles.  As the material approaches the white
dwarf, a shock front is created where impact energy is released in the
form of extreme ultraviolet (EUV) and X-ray photons.  Surface heating,
of up to a few hundred thousand degrees Kelvin, and possible
penetration of the surface by material blobs completes the accretion
region production of high energy flux.
\citet{cropper1990} presents a detailed review of polars.

Since the accretion regions in polars are the site of the majority of
high energy emission, we would expect that EUV and X-ray observations
would provide a wealth of information about their accretion
geometry. This is indeed the case as shown, for example, by
\citet{sirk1998}.  Differentiation and study of eclipses of the
accretion region by the secondary, the far and near accretion stream,
and the white dwarf (a self-eclipse; when the rotation of the CV
system causes the accretion pole to pass behind the limb of the white
dwarf), allows a number of system parameters to be determined. These
include the inclination of the system, the position of the accretion
pole on the white dwarf, and if the white dwarf is accreting at one or
both of its magnetic poles.  With this type of study in mind, one of
us (SBH) placed QQ~Vul on the target list of the {\it EUVE} satellite
Right Angle Program \citep{mcdonald1994}.

\section{Previous Observations of QQ Vul}
Serendipitously discovered in a survey of soft X-ray sources
\citep{nugent1983}, QQ~Vulpeculae was confirmed as an AM Her binary 
through detection of circular and linear polarization
\citep{nousek1982,nousek1984}.  QQ~Vul has a relatively long orbital
period for a polar, $P_{orb}=222.5$ min \citep{nousek1984}, and
although the strength of the magnetic field has not been directly
measured, it is assumed to be fairly typical, $\sim20-30$ MG
\citep{liebert1985}.  Blackbody fits to the spectrum of QQ~Vul yield a
temperature of $T\sim2\times10^{5}$ K for the X-ray heated accretion
regions \citep{nousek1984}. Mass estimates of M$_1$=0.58M$_{\odot}$
and M$_2$=0.35M$_{\odot}$ have been determined for QQ~Vul by
\citet{mukai1987}.

The initial multi-wavelength observations of QQ~Vul \citep{nousek1984}
were able to place some constraints on the system geometry.  These
observations suggested a system with a magnetic pole tilted
$\sim75^o-85^o$ from our line of sight during the linear polarization
pulse peak, an orbital inclination of $46^o < i < 74^o$, and a stellar
latitude of the accreting magnetic pole in the range $63^o < \Delta <
80^o$.  Circular and linear polarization observations
\citep{nousek1984} have revealed a weak and diffuse linear
polarization pulse centered on maximum light, indicating that the near
field accretion column is always in sight, although the pole does
graze the limb during a self-eclipse \citep{cropper1998}.  The
polarization data also suggest that there is non-radial accretion flow
\citep{mccarthy1986,cropper1998} implying a ``kink" in
the accretion stream which flows to the magnetic pole.

Throughout its history of observations, QQ~Vul has repeatedly shown a
complex and varying X-ray light curve.  Studies undertaken with the
use of {\it Einstein} \citep{nousek1984}, {\it ROSAT} \citep{beardmore1995},
and {\it EXOSAT} \citep{osborne1986,osborne1987} have all shown the
complexities apparent in the X-ray component of QQ~Vul.
\citet{osborne1987} found that the soft X-ray count rate had doubled
within a period of two years and that the shape of the light curve
they observed was indeed quite different from the initial X-ray light
curve of QQ~Vul \citep{nousek1984}.  Figure 1 in
\citet{osborne1987} provides a comparison of the different X-ray light
curves previously obtained for QQ~Vul.

There has been conflicting evidence for whether or not QQ~Vul
possesses two accreting poles.  While the double peaked nature of the
soft X-ray light curves of \citet{osborne1986,osborne1987} might lead
one to interpret it as two-pole accretion, it is noted there that the
second pole is not evident in the optical light curve.  Other
observations \citep{beardmore1995} detected soft X-ray spectral
variations as a function of orbital phase which could be modeled by an
extended multi-temperature accreting region or by two accreting poles
with slightly different temperatures.  However, all of these previous
observations do agree that if two-pole accretion is taking place, the
primary accreting pole is a weaker source in soft X-rays.  Recent
polarization data seem to strongly suggest that QQ~Vul is undergoing
two-pole accretion.  Optical polarization data from
\citet{schwope1991} cannot be explained by one-pole accretion and
a second linear polarization peak, seen in the data of \citet{cropper1998},
requires a second accreting region to be in view at certain binary 
phases.

\section{{\it EUVE} Observations and Data Analysis}\label{data}
The EUV photometric data were obtained with the {\it EUVE} satellite
using the right-angle pointing Scanner Telescopes A and B.  Scanner A
imaged QQ~Vul through a Lexan/Boron filter ($\lambda_{\rm peak} =
89{\rm \AA}$), sensitive in the bandpass $50-180{\rm \AA}$ while
Scanner B data were obtained in an Al/Ti/C filter ($\lambda_{\rm peak}
= 171{\rm \AA}$), sensitive in the bandpass $160-240{\rm \AA}$.
Details of the photometric properties of the imaging telescopes on
board the {\it EUVE} may be found in \citet{sirk1997}.

Our observations of QQ Vul began on 1996 Aug 11 (GMT) and continued
through 1996 Aug 16 (GMT), spanning $\sim1.5\times10^5$s, or a total
of $\sim11P_{orb}$.  The data were passed through {\it EUVE} standard
processing and delivered to us in compressed format on CD-ROM.  We
then extracted the scanner observations using the standard {\it EUVE}
data analysis software packages within IRAF.  Photometry was performed
using an aperture with a seven pixel radius centered on the
coordinates of the source and a background annulus having a radius of
twenty pixels also centered on the object.  Due to the large
difference in signal-to-noise obtained in the two data sets, every 100
data points (photon events) were binned together for the raw data from
Scanner A, and every 30 data points were binned together for the raw
data from Scanner B.  An IDL program (written by M. Sirk) was
then used to produce light curve data files that were phased according
to the ephemeris of the inferior conjunction of the secondary star in
QQ~Vul, HJD 2448446.4710(5) + 0.15452011(11)E
\citep{schwope1998}.  Finally, our resultant light curves, in both
Scanner A (Lexan/B) and Scanner B (Al/Ti/C), were rebinned to 0.005 in
phase.

Figures \ref{sca_lcurve} and \ref{scb_lcurve} present our obtained
{\it EUVE} light curves phased on the \citet{schwope1998} ephemeris.
Convolving the mean EUV count rate of 0.01 counts/second with the
effective area as a function of wavelength for the Lexan/B filter
\citep{sirk1999}, we find that the observed EUV flux is
$4.74\times10^{-14}$ ergs$\;$s$^{-1}$cm$^{-2}$.  It is
interesting to note that Scanner B, viewing QQ~Vul through the Al/Ti/C
filter, detected anything at all.  At a wavelength of $171{\rm \AA}$,
and a hydrogen column density of $N_H\sim10^{20}{\rm cm^{-2}}$
\citep{osborne1986} for a distance to QQ~Vul of 215 pc
\citep{mukai1986}, the optical depth is $\tau\sim14$.  At such a large
optical depth, the ISM transmission of photons from QQ~Vul at this
wavelength is essentially zero.  We would therefore not expect to
detect photons through this filter and indeed no other polar has been
detected by {\it EUVE} in Scanner B at these wavelengths.  However,
the Al/Ti/C filter is known to have an X-ray leak peaking near 44${\rm
\AA}$
\citep{finley1988,vallerga2000}, and we conclude that the data
collected here with Scanner B is an X-ray light curve for QQ~Vul with
a mean effective wavelength of $\sim$44${\rm \AA}$.  We note that this
is not the first detection of an X-ray leak with the Al/Ti/C filter;
X-ray leaks were also reported in {\it EUVE} observations of the nova
V1974 Cygni \citep{mac1996,string1996}.

One flaw in our conclusion concerning the X-ray light curve would be
if QQ~Vul were actually quite close by in space, say less than 50 pc.
We therefore independently re-determined the distance to QQ Vul using
Bailey's method \citep{bailey1982} and newly obtained infrared
observations.  Bailey's method relies on the relationship between
certain physical parameters of the secondary star in the CV system and
the distance to that system. The relation is:
\begin{equation}
{\rm log}\,d = \frac{K}{5} + 1 - \frac{S_K}{5} + 
{\rm log}\,\frac{R_2}{R_{\odot}}
\end{equation}
where $d$ is the distance, $K$ is the $K$-band magnitude of the
secondary star, $S_K$ is the $K$-band surface brightness of the
secondary, and $R_2$ is the radius of the secondary star.

Using data kindly obtained by M. Huber with the Wyoming Infrared
Observatory on 1998 Aug 30 UT (9:30 hours), we find that QQ~Vul had a
$K$ magnitude of 14.0$\pm$0.1 mag.  Taking $S_K=4.5$
\citep{bailey1982}, and $R_2=0.43 R_{\odot}$ \citep{nousek1984}, the
distance to QQ Vul is determined to be $342$ pc; a value which is in
agreement with earlier measurements which suggest a lower limit of
$\sim215$ pc \citep{mukai1986}.  $V, R,$ and $I$ observations of
QQ~Vul by \citet{mukai1988} detected a visual ``companion star" to
QQ~Vul with $K\sim14.5$ (obtained using $V-K$ colors derived for $K$
spectral type stars).  If possible contamination from this star in the
infrared (i.e., $K$ band) is taken into consideration, QQ~Vul would be
even farther away.  Therefore, it seems highly unlikely that the
detected signal in Scanner B is due to $160-240{\rm \AA}$ photons but
is instead the aforementioned X-ray leak.

The effective bandpass of the X-ray leak in the Al/Ti/C filter is
roughly triangular in shape and covers the range of $15{\rm \AA}<
\lambda < 68 {\rm \AA}$. The peak throughput, at 68${\rm \AA}$, is
$\la$2\% of the normal filter transmission near 171${\rm \AA}$ and has
zero sensitivity to photons with a wavelength below $15{\rm \AA}$
\citep{sirk1999}.  Using the effective area ratio of the Lexan/B
filter to the Al/Ti/C filter \citep{sirk1999}, and the fact discussed
above concerning the total absence of long wavelength photons, we can
determine that the X-rays observed for QQ~Vul have wavelengths
from $15-68{\rm \AA}$, with a mean effective central wavelength near
44${\rm \AA}$. Performing an approximate integration under the
triangular bandpass and convolving it with the effective area as a
function of wavelength, we find the observed X-ray flux to be
$3.60\times10^{-10}$ ergs$\;$s$^{-1}$cm$^{-2}$.  X-ray
fluxes ranging from $1.5\times10^{-12}$ ergs$\;$s$^{-1}$cm$^{-2}$
\citep{osborne1987} to $2.25\times10^{-11}$ ergs$\;$s$^{-1}$cm$^{-2}$ 
\citep{beardmore1995} have been reported in previous studies, which, 
along with the value determined here, reflect the variability of the
source.

We have thus obtained simultaneous time-resolved photometric data in
the X-ray ($\lambda\sim44{\rm \AA}$) (Figure~\ref{scb_lcurve}) and EUV
($\lambda_{\rm peak}=89{\rm\AA}$) (Figure~\ref{sca_lcurve}) wavelength
regions which allow us to make a direct comparison of the emitting
character of this system in these two wavelength regimes.

\section{Discussion and Conclusions}\label{discussion}
\begin{figure}
\centerline{
\psfig{figure=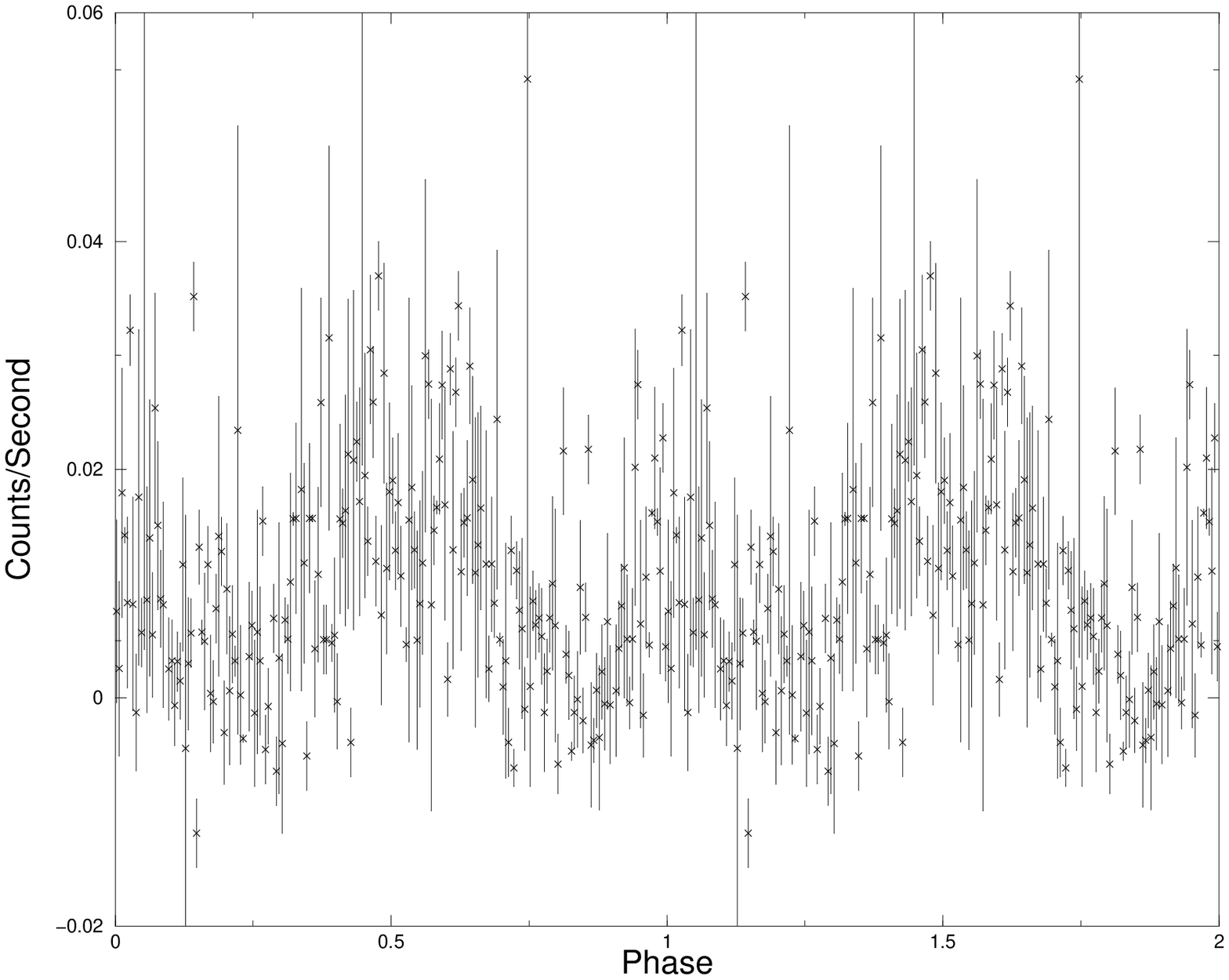,width=5in,angle=270}
}
\caption[]{EUV light curve for QQ Vul observed by {\it EUVE}/Scanner
A with the Lexan/B filter ($\lambda_{\rm peak} = 89{\rm \AA}$).  The
data are folded on the orbital period of 222.5 min according to the
ephemeris of \citet{schwope1998}.  Each point represents 0.005 in
phase and the data cover nearly 11 consecutive binary orbits of
QQ~Vul. The error bars are 1$\sigma$ values in the count rate for each
phase bin. \label{sca_lcurve}}
\end{figure}

\begin{figure}
\centerline{
\psfig{figure=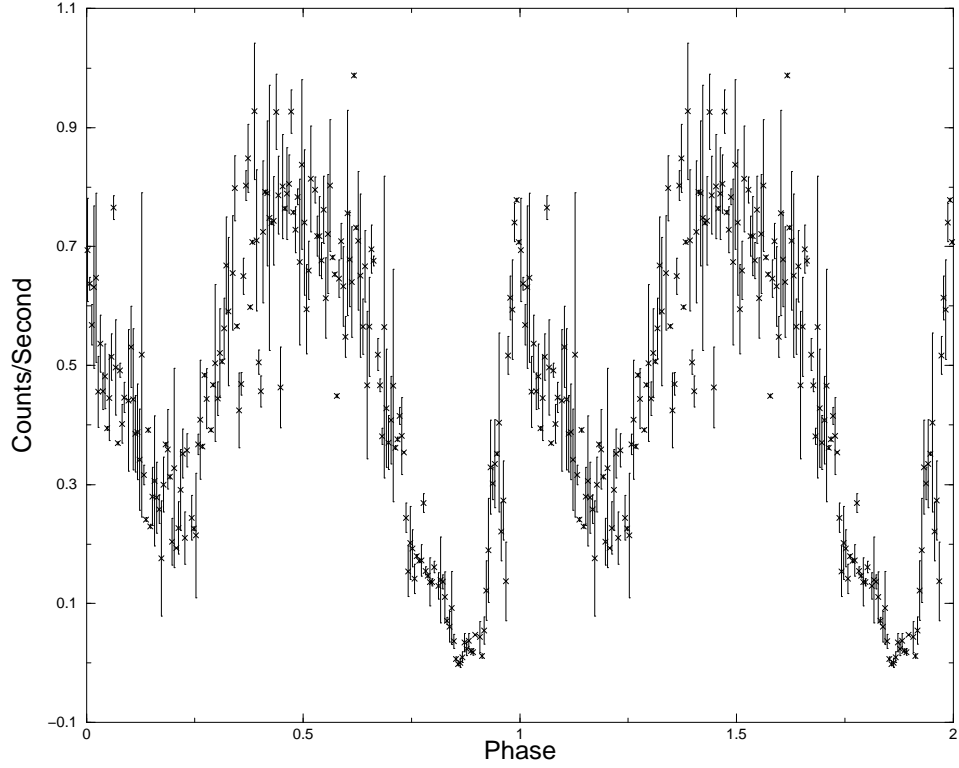,width=5in,angle=270}
}
\caption[]{X-ray light curve for QQ Vul observed by {\it EUVE}/Scanner B
with the Al/Ti/C filter.  These data represent an X-ray leak in the
Al/Ti/C filter and have a mean effective wavelength near $44{\rm
\AA}$. This is the first X-ray detection of a polar with the {\it
EUVE} satellite.  The data are folded on the orbital period of 222.5
min according to the ephemeris of \citet{schwope1998}.  Each point
represents 0.005 in phase and the data cover nearly 11 consecutive
binary orbits of QQ~Vul. The error bars are 1$\sigma$ values in the
count rate for each phase bin. \label{scb_lcurve}}
\end{figure}

The EUV and X-ray light curves, Figures \ref{sca_lcurve} and
\ref{scb_lcurve} respectively, both show a double peak shape with
minima occurring near phase 0.2 and phase 0.85 and maxima at
phases 0.0 and 0.45.  While each light curve reveals similar gross
features, we note that there is far less change and detail in the EUV
data. This could be due to the fact that QQ~Vul has a broader, more
diffuse EUV emitting region but a smaller, better defined X-ray
emitting region.  The modulations of both light curves are uneven in
their minima and maxima. The minima alternate between a deep,
essentially complete eclipse at phase 0.85 to a less deep and
well-defined dip near phase 0.2, while the maxima shift between a
narrow peaked one near phase 0.0 to a brighter, broader one covering
about 0.4 in phase, centered at 0.45.  The two maxima, while showing
that the secondary pole is stronger in intensity, are probably nearly
equal in phase extent and overall shape, the narrower one being
``cut-off" around phase 0.9 by an eclipse of the magnetic pole
accretion region by the near-field accretion stream \citep{sirk1998}.

Interpreted as a two-pole accretor, the locations of the two poles
would have centers near binary phases 0.1 and 0.55, that is, almost
directly along the line of centers of the binary.  The eclipse by the
near-field stream of the accretion region facing the secondary star
occurs before phase 0 as is the case in most polars
\citep{sirk1998}. The magnetic pole on the far side of the white dwarf
suffers no eclipse, thus, it is visible in phase for approximately
one-half of the orbital period, and its shape is consistent with a
spot latitude of $55^o-75^o$ \citep{sirk1998} given a binary
inclination of $60^o-90^o$ (see below).
 
Figure \ref{zoom} is a close-up of the X-ray light curve minimum near
phase 0.8.  It appears that this minima has two components.  The
first half of the broad dip, starting at phase 0.7, has a slow decline
up to phase 0.85 and is likely to be the result of an eclipse by the
near-field accretion stream. The remaining part of this dip shows an
abrupt drop to near zero counts and appears to be flat bottomed from
phase 0.85 to 0.92, with the most likely cause being a stellar eclipse
of the X-ray emitting region by the secondary star.  If true, this 
constrains the system inclination to be greater than $60^o$.

An interesting feature appearing in the X-ray light curve (Figures
\ref{scb_lcurve} and \ref{zoom}), but not seen in the EUV light curve,
is the narrow dip which occurs at phase 0.96. Using other polar
light curves as a guide, this narrow dip feature is likely to
correspond to an eclipse of the accretion region by the far-field
accretion stream. It may also be present in the EUV data but the noise
level precludes its discovery.  While unresolved, the short duration
of this narrow feature (0.015 in phase or 3 min) provides strong
evidence for the compactness of the hard X-ray emitting region in
QQ~Vul. Translating this time in to a size on the white dwarf surface
(without correction for latitude and assuming R$_{WD}$ = 7000km) we
find an emitting region diameter of 660 km or, if circular, $f$=0.002.

Taking the gross ratio of the low EUV count rate to the higher X-ray
count rate (even as a leaked signal) one could conclude that the
magnetic field strength in QQ~Vul is relatively low, possibly less
than 10-30 MG \citep{ramsay1994}.  However, while in general a large
X-ray to EUV ratio indicates a lower magnetic field strength in
polars, this is not always the case with the difference attributed to
the structure and size of the accretion region \citep{sirk1998}.  

Figure \ref{hratio} presents the hardness ratio (X-ray/EUV) for
QQ~Vul. Due to the low value of the flux in the EUV light curve, both
the X-ray and EUV light curves were re-binned to 0.05 in phase,
thereby allowing the hardness ratio to not be dominated by noise
spikes due to the low EUV flux values.  Figure \ref{hratio} exhibits 
an increased hardness near phase
0.25, with a sharp rise near phase 0.3.  This peak might indicate
the emergence of the second accreting pole.

A comparison of our QQ~Vul X-ray light curve with previous high energy
observations shows that the continuous orbital variations and temporal 
changes, noted by \citet{osborne1987}, appear to be persistent.  Some
similarities, however, do exist between our X-ray light curve and the
1983 Oct and 1985 Jun {\it EXOSAT} light curves discussed in
\citet{osborne1986,osborne1987}.  For example, the unequal minima and
maxima and even the presence of a narrow dip, probably due to an
eclipse of the accretion region by the far-field accretion stream.
This narrow dip feature occurs near phase 0.03 in the 1983 Oct {\it EXOSAT}
light curve and 0.08 in the 1985 Jun {\it EXOSAT} light curve,
compared with phase 0.96 seen in our light curve [according to the
ephemeris of \citet{schwope1998}].  The 1985 Sep {\it EXOSAT} light
curve \citep{osborne1987}, is very different from our present data as
it exhibits a much different shape with nearly equal maxima and yet
again a narrow dip but one which appears at yet another phase (phase
0.71) within the light curve.  The fact that the narrow dip is always
present but changes phase indicates a movement, within the binary, of
the far-field accretion stream similar to that observed in HU Aqr
\citep{schwope1998a}.

\acknowledgments
KEB is supported by a graduate assistantship from the University of
Wyoming.  SBH acknowledges partial support for this work from NASA
cooperative agreement NCC5-138 through an {\it EUVE} Guest Observer
mini-grant and from NASA ADP grant NAG5-4233.  The authors wish to
thank Jennifer Cash and David Ciardi for their assistance with data
reduction, Martin Sirk for his assistance with data reduction and for
extremely useful discussions concerning the performance of the {\it
EUVE} filters, Mark Huber for providing $K$-band photometry for
QQ~Vul, and Axel Schwope for supplying us with his QQ~Vul ephemeris
prior to publication.

\newpage

\begin{figure}
\centerline{
\psfig{figure=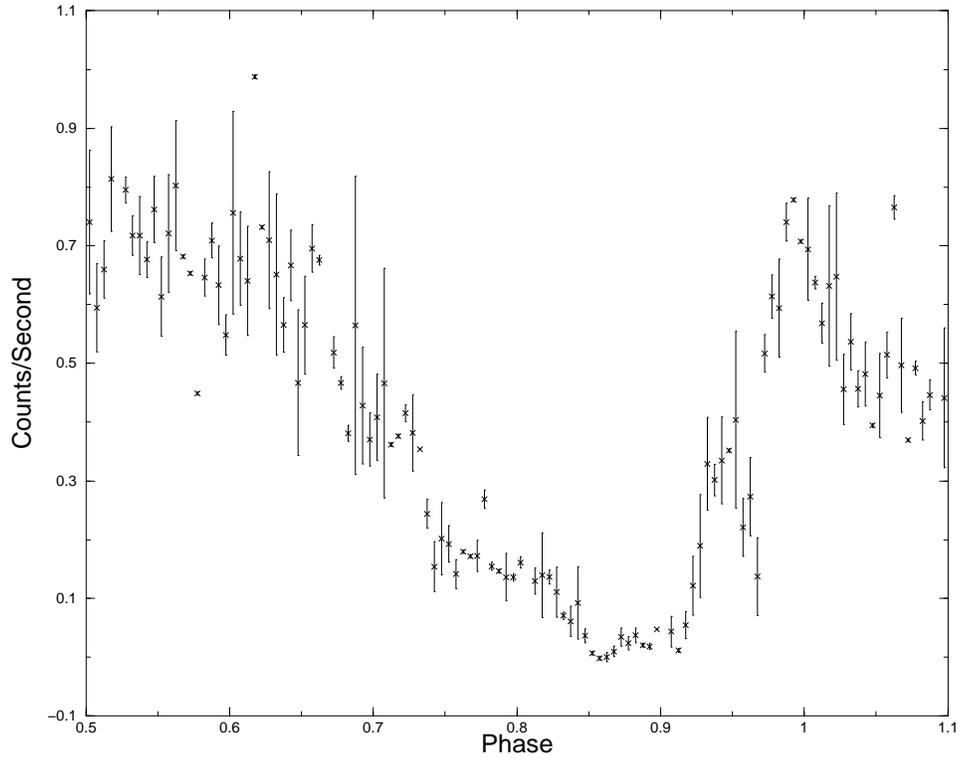,width=5in,angle=270}
}
\caption[]{An expanded view of the X-ray minimum in QQ~Vul
at phase 0.8.  One can clearly see the structure in the minimum
including the near-field accretion stream absorption and the probable
stellar eclipse of the X-ray emitting region by the secondary star
(phases 0.85 to 0.92).  The narrow dip seen near phase 0.96 is caused
by an eclipse of the X-ray emitting region by the far-field accretion
stream. \label{zoom}}
\end{figure}

\begin{figure}
\centerline{
\psfig{figure=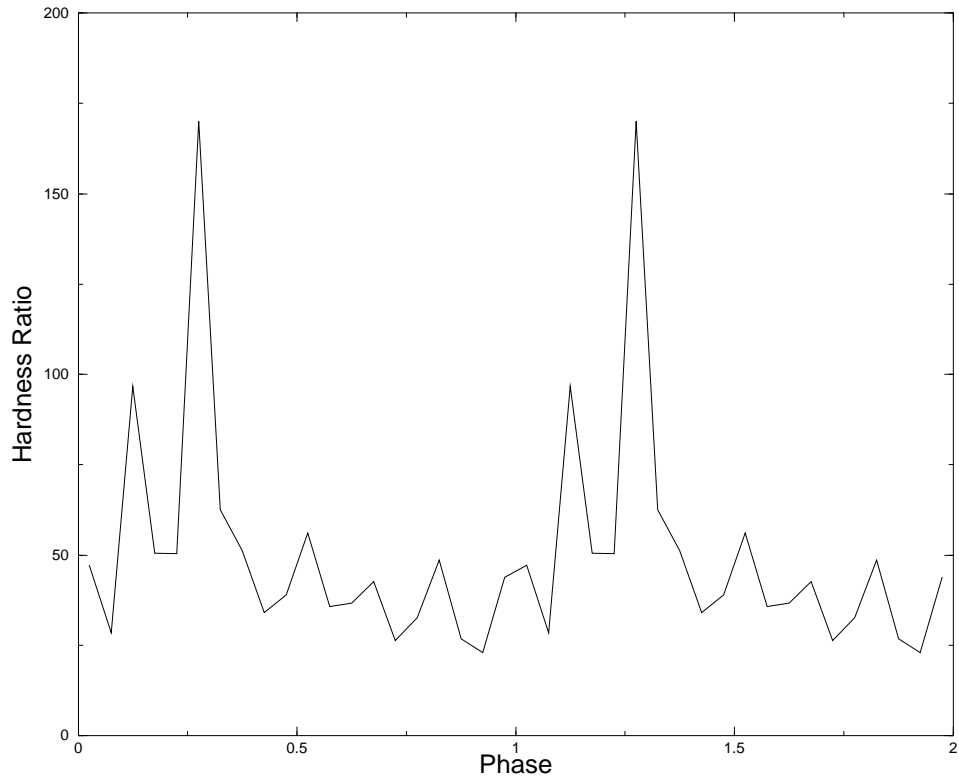,width=5in,angle=270}
}
\caption[]{The hardness ratio ($\lambda\sim44{\rm \AA} /
\lambda\sim89{\rm \AA}$) for QQ~Vul.  The X-ray and EUV light curves 
were re-binned to 0.05 in phase for this plot.  The ratio shows a peak
near phase 0.3, possibly associated with the appearance of the second
accreting pole. \label{hratio} }
\end{figure}

\end{document}